# Delta-oriented Architectural Variability Using MontiCore

Arne Haber
Software Engineering
RWTH Aachen University,
Germany
http://www.se-rwth.de/

Thomas Kutz
Software Engineering
RWTH Aachen University,
Germany
http://www.se-rwth.de/

Holger Rendel
Software Engineering
RWTH Aachen University,
Germany
http://www.se-rwth.de/

Bernhard Rumpe
Software Engineering
RWTH Aachen University,
Germany
http://www.se-rwth.de/

Ina Schaefer
Institute for Software Systems
Engineering
TU Braunschweig, Germany
http://www.tu-bs.de/sse

## ABSTRACT

Modeling of software architectures is a fundamental part of software development processes. Reuse of software components and early analysis of software topologies allow the reduction of development costs and increases software quality. Integrating variability modeling concepts into architecture description languages (ADLs) is essential for the development of diverse software systems with high demands on software quality. In this paper, we present the integration of delta modeling into the existing ADL MontiArc. Delta modeling is a language-independent variability modeling approach supporting proactive, reactive and extractive product line development. We show how $\Delta$-MontiArc, a language for explicit modeling of architectural variability based on delta modeling, is implemented as domain-specific language (DSL) using the DSL development framework MontiCore. We also demonstrate how MontiCore's language reuse mechanisms provide efficient means to derive an implementation of $\Delta$-MontiArc tool implementation. We evaluate $\Delta$-MontiArc by comparing it with annotative variability modeling.

## Categories and Subject Descriptors

D.2.11 [**Software Engineering**]: Software Architectures—*Languages*; D.2.2 [**Software Engineering**]: Design Tools and Techniques—*Computer-aided software engineering (CASE)*

## Keywords

Software Architectures; Delta Modeling; DSL Development; DSL Reuse

## 1. INTRODUCTION

In modern software development processes, the modeling of the system architecture is an integral part [26]. Architectural description languages (*ADL*) provide modeling elements to describe system components, their ports as interaction points, and the communication connections between components [25]. The structuring of the system into small and manageable parts decreases complexity of single development tasks and allows reuse of previously developed components. In addition, the analysis of non-functional system properties is possible by considering component topologies and communication paths [9].

Modern software systems exist in many different variants in order to adapt to changing user requirements or application contexts, while imposing high demands on system quality and correctness [33]. In order to develop a set of diverse systems with the desired quality attributes, the envisaged system variability has to be represented explicitly on the architectural level. Variability modeling concepts have to support the modular representation of variability in order to enable a scalable modeling of architectural variability, while still being flexible and expressive enough to capture differences between different system variants.

Delta modeling [4, 32, 31] is a language-independent variability modeling concept that supports the flexible modular modeling of variability for proactive, reactive and extractive product line development [23]. In delta modeling, a set of systems is described by a designated core system and a set of system deltas that specify modifications of the core product to obtain other products. A particular system variant is obtained by applying the modifications of a selected subset of the deltas to the core product. In previous work [14], we presented the first application of delta modeling to an ADL and proposed a proof-of-concept language to represent architectural variability. This language only contained basic operations to add and remove architectural elements and modify components by changing their internal structure.

In this paper, we present $\Delta$-MontiArc, a delta-oriented variability modeling language explicitly designed to represent architectural variability. In addition to the basic modification operations of delta modeling that were already presented in [14], $\Delta$-MontiArc contains designated operations specifically tailored to model architectural variability, such as replacing compatible components or re-wiring component connections. In order to resolve conflicts between modifi-cations targeting the same architectural modeling element, e.g., the same component, an ordering constraint attached to each delta determines which other deltas have to be or must not be applied before the respective delta. The selection of the applicable deltas for a particular system variant is specified separately from the deltas in order to support configurability by external product configuration tools. We implement Δ-MontiArc as a domain-specific language (DSL) using the DSL framework MontiCore [12]. We extend the existing ADL MontiArc [13] developed with MontiCore by relying on MontiCore's language reuse concepts. In order to evaluate the potential for delta modeling of software architectures, we compare Δ-MontiArc with annotative variability modeling approaches using two case studies from the embedded system domain. The results show that the delta models has a clearer and more understandable structure.

This paper is structured as follows: In Section 2, the ADL MontiArc and the DSL framework MontiCore are described. Section 3 gives an introduction to Δ-MontiArc. Its implementation with MontiCore is presented in Section 4. Product generation from Δ-MontiArc-models is explained in Section 5. Section 6 describes case studies. Related work is discussed in Section 7. Section 8 summarizes the paper.

## 2. THE ARCHITECTURAL BASIC LANGUAGE

The delta-oriented variability modeling approach for software architectures presented in this paper is based on the existing architectural description language MontiArc [13] which is developed using DSL-framework MontiCore [12]. We chose MontiArc because it provides all interesting concepts an ADL needs. MontiArc and MontiCore are introduced in the following.

### 2.1 MontiArc

Architectural description languages (*ADLs*) are modeling languages that support design and development of (software) systems. Therefore, they offer modeling elements like components, interfaces, and connectors, that allow a high-level description of a system [25]. Based on this abstract model that contains the most important system parts, their relation, and interaction, reasoning about structural, as well as non-functional, properties is possible in an early development stage [9]. MontiArc [13] is a textual ADL tailored to distributed information-flow architectures and their message-based asynchronous communication. Following taxonomy for ADLs presented in [25], MontiArc components are units of computation or storage. As components are only accessed and connected by their defined interfaces, their interior implementation is encapsulated.

As an example, we present the logical architecture of a multicopter control system. These helicopters with four or more rotors are mostly unmanned and may be extended with certain sensors and optics to improve their flight-behavior and functionality. Figure 1 contains the definition of component `FlightController`. Its interface description is given by incoming and outgoing ports that have a type (e.g. `SteeringCmd`) and a name (e.g. `engine1`). The component realization is given by an inner decomposition to subcomponents. These are component instances of a certain type (e.g. `GyroEval`) and a name (e.g. `gEval`). In MontiArc naming of subcomponents and ports is optional, as long as a used type is unique in a component definition. If no name is explicitly given, a subcomponent, respectively a port, is accessible by its type's name. Subcomponents and the outer interface are connected by ports that capture a unidirectional communication. The component depicted in Figure 1 receives several signals. Data type `SteeringCmd` encapsulates the typical steering commands *nick*, *roll*, *yaw* and *throttle*. The `SteeringMode` is a flag which defines whether accelerometer or heading-hold flight mode is used. In accelerometer flight mode, the sticks of the transmitter control the angle of the multicopter whereas in heading hold mode the angular velocity is controlled. Furthermore, the `FlightController` component receives signals from the gyroscope and the accelerometer sensors. These signals are converted to float arrays by the subcomponents `gEval` and `AccEval`. The signals `SteeringCmd`, `SteeringMode`, and the converted arrays are processed by a `SteeringCmdProcessor` that calculates the power for each rotor engine. This output is post-processed by an `outputProcessor` that transforms the values into signals understandable by the rotor engine hardware.

Listing 1 contains the textual representation of the example depicted graphically in Figure 1 in MontiArc. A component definition starts with the keyword `component` followed by its type name (l. 1). A components body contains arbitrary many architectural elements: ports that describe component interfaces, subcomponents as instantiation of other component definitions, inner components as private component definitions, and connectors to capture communication. The interface of a component is defined using keyword `port` followed by incoming (ll. 5–8) and outgoing (ll. 9–12) ports. A port always has a type (e.g. `SteeringCmd` l. 5) and an optional name. Subcomponents are instantiated using the keyword `component` followed by a component type and an optional name (ll. 14–17). Some components need parameters to be instantiated. For example component `SteeringCmdProcessor` (l. 14) is instantiated as subcomponent `scp` using parameter value 4. This way internal algorithms are optimized to handle four rotors. Connectors between subcomponents and the outer interface or other subcomponents are inserted automatically using the keyword `autoconnect` (l. 2). Parameterized with parameter `port` connections between ports with the same name and a compatible type are automatically derived. An alternative parameter is `type` that creates connections based on the port types only. Additionally, connections may be created manually, if preferred or required.

### 2.2 Implementation of MontiArc

Language design does not only comprise the definition of a language itself, but also the associated tool support. The DSL-framework MontiCore [12] supports the definition and generation of all relevant language processing artifacts for a specific language that is specified by a grammar similar to EBNF. Generated artifacts are among others the abstract and concrete syntax of a language, a parser, a lexer, and a set of runtime components. Additionally, support for creation of symboltables and automated checking of context conditions is provided. MontiCore allows defining language processing workflows which transform input artifacts into output artifacts based on an abstract syntax representation. Thus, domain-specific languages and the accompanying tool support can be developed in an efficient way.

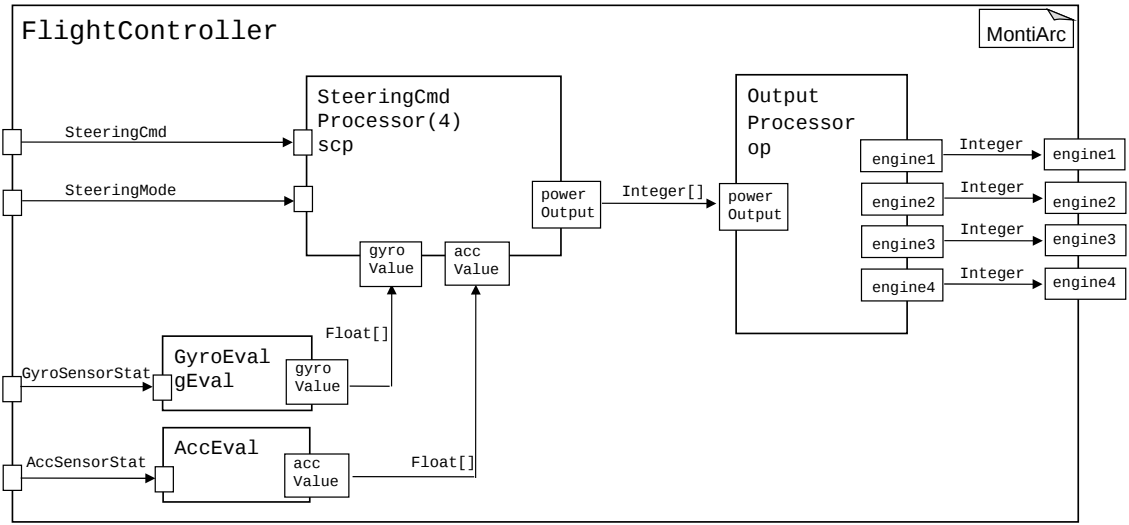

Figure 1: **MontiArc** architecture of a quad-copter control system

```
component FlightController {
  autoconnect port;

  port
    in SteeringCmd,
    in SteeringMode,
    in GyroSensorStat,
    in AccSensorStat,
    out Integer engine1,
    out Integer engine2,
    out Integer engine3,
    out Integer engine4;

  component SteeringCmdProcessor(4) scp;
  component OutputProcessor op;
  component GyroEval gEval;
  component AccEval;
}
```

Listing 1: Component `FlightController`

MontiCore additionally offers means for language extension and reuse by inheritance and embedding. Both concepts simplify the implementation of architectural variability in MontiArc. Language inheritance means that an existing grammar for a particular language can be extended and refined by defining new grammar rules or redefining existing rules. Inheritance is also used for extending generated abstract syntax representation which is implemented by Java classes. Thus, existing language processing workflows which use these Java classes do not need to be rewritten. The concept of language embedding allows defining a grammar with gaps which can be filled later with other domain-specific languages. For example, a grammar can be defined with a gap for a constraint language where boolean logic or OCL can be embedded in a further step. MontiCore does not only facilitate the combination of abstract syntax by means of language embeddings, but allows also to reuse the other parts of a language implementation, e.g. symboltables and context conditions.

The ADL MontiArc described above is developed using MontiCore. The tool support provided for MontiArc contains artifacts such as a symboltable, a context condition checker, or a code generator. These tools process the abstract syntax of MontiArc models. Therefore, they may be reused in extensions of MontiArc as well. The MontiArc ADL can be extended by using the interface `ArcElement` as an extension point on the grammar level. By implementing this interface in a sub-grammar, new architectural elements may be added to component definitions. Existing tools may be reused. For this purpose, sub-language tools only have to process new elements defined by the sub-language. Original MontiArc elements are handled by the existing MonitArc tool implementation.

## 3. VARIABILITY IN ARCHITECTURAL DESCRIPTIONS

$\Delta$-MontiArc represents the variability of software architectures based on delta modeling, a language-independent variability modeling approach. While in [14] only the conceptual feasibility of this idea was demonstrated, in this section, we present a refined language which comprises linguistic constructs specifically tailored for representing architectural variability. The variability of a software architecture in $\Delta$-MontiArc is represented by a specific core architecture that is described by a set of core models in MontiArc. The core architecture is a complete architecture description of one product. Additionally, in $\Delta$-MontiArc, a set of delta models can be specified. Each delta model contains modifications of MontiArc models to obtain architectures for further products. In order to generate the architecture description (given by a MontiArc model) for a particular product, a subset of the delta models is determined, and the contained modifications are applied to the core models. In the following, we present the different modification operations that can be specified in delta models, define how architectures for particular products can be derived and illustrate the approach using the multicopter example.

*Modification operations.*
$\Delta$-MontiArc provides a number of statements for modi-

| Statement | Affected elements |
|---|---|
| `modify component` | components |
| `add` / `remove` | ports, components, parameters |
| `connect` / `disconnect` | connectors |
| `rename` | ports, components, parameters |
| `replace component` | components |
| `expand autoconnect` | autoconnect |
| `introduce autoconnect` | autoconnect |
| `remove unreachables` | ports, components, connectors |

Table 1: Overview of Modification Operations

fication of a architectural models defined in MontiArc [13]. An overview of modification operations is given in Table 1. The modification of an architectural component is specified using the keywords `modify component` followed by a list of modification operations in curly brackets. By using the `add` operation, ports, subcomponents, configuration parameters, and autoconnect statements can be added to a component. The inverse `remove` operation allows to remove these kinds of elements. Ports are connected with each other using the `connect` operator. In order to remove a connector, the `disconnect` statement can be used analogously.

These basic modification operations are sufficient for defining all possible modifications of MontiArc architecture models. However, some modifications are difficult to describe when using only `add`, `remove`, `connect`, and `disconnect` operations. For example, renaming a port does not only require to remove the old port and to add the new one, but it also requires to update all connectors that used the old port name. This results in a number of modification operations which can be error-prone, if it has to be done manually. For this reason, we define further modification operations to alleviate the modeling of architectural variability.

The `rename` operator allows renaming ports, subcomponents, and configuration parameters such that also all affected elements are updated. This means, that after renaming a port $p$ in component $c$, also all connectors which use $p$ as source or target are updated with the new port name. To achieve that, all core models have to be examined, since $p$ can not only be used in the connectors of $c$, but also in the connectors of components where $c$ is used as a subcomponent. Similarly, renaming a subcomponent $sc$ of component $c$ causes an update of all connectors in $c$ which use a port of $sc$ in their source or target. When renaming a configuration parameter $cp$ in component $c$, also the configurations of subcomponents which use $cp$ as a variable are updated.

The `replace` operator can be used to substitute a subcomponent $sc_1$ of component $c$ with another subcomponent $sc_2$, if the interfaces of $sc_1$ and $sc_2$ are compatible. Using this operator does not only save the corresponding `remove` and `add` statements but also the re-wiring of all connectors that used the replaced subcomponent. The interfaces of $sc_1$ and $sc_2$ are compatible, if first the number of incoming ports is the same and there is a mapping between the two sets of incoming ports such that for each incoming port $p$ in $sc_1$ there is an incoming port $p'$ in $sc_2$ which is of the same type or of a supertype of $p$. Second, the number of outgoing ports in $sc_2$ is at least the number of outgoing ports in $sc_1$ and the outgoing ports of $sc_2$ are of the same or of a subtype of the outgoing ports in $sc_1$.

As mentioned in Section 2.1, some components need to be parametrized when instantiated as subcomponents. This configuration can be adjusted with the `modify component` statement followed by an arbitrary number of parameter assignments in parentheses. In this way new values can be assigned to existing configuration parameters which are explicitly mentioned to improve readability.

The `autoconnect` statement in MontiArc automatically connects compatible ports depending on the used `autoconnect` configuration. Ports are connected automatically based on their types (`autoconnect type`) or their names (`autoconnect port`). This statement can also be used in a delta model to overwrite the configuration of the core component. Disabling automatic connections can be achieved with the `autoconnect off` statement. In addition, in Δ-MontiArc advanced statements are defined for handling the autoconnect concept. The `expand autoconnect` operation makes all implicitly defined connectors explicit. These connectors can then be removed in subsequent deltas by the `disconnect` statement. The `expand autoconnect` operation also disables the `autoconnect` configuration for the currently modified component. Hence implicit connectors do not exist anymore, but the modeled system is not affected directly. The `introduce autoconnect` statement works in the opposite way. All connectors which can be recreated automatically are removed, and the corresponding `autoconnect` configuration is set. This way previously unconnected ports with the same name are connected as well.

The operator `remove unreachables` removes all ports and subcomponents which do not contribute to the output of the currently modified component. In doing so, in a first phase all subcomponents which do not contain any outgoing connectors are removed together with all their incoming connectors. Since this step can result in new subcomponents without outgoing connectors, it is recursively repeated for all affected components. In a second phase, all ports which are no more used by a connector are removed.

The operators `expand autoconnect`, `introduce autoconnect` and `remove unreachables` can be used either globally or locally. When defined globally (i.e., outside of a concrete component), the modifications are applied to all architectural core models, whereas a local definition only affects the currently modified component.

*Product configuration.*
Compared to the previous version of Δ-MontiArc [14], the delta models in the language presented in this paper are independent of product features specified in feature models. In detail, this means that the application condition as a logical expression over features is no longer defined within the delta model. This makes the modifications specified in the delta models more reusable, e.g., for a different product line for the same application domain.

However, Δ-MontiArc allows the definition of a logical expression over delta models associated with each delta

```
1  deltaconfig DeltaWolf {
2    PressureSensor,
3    HeightHold,
4    HexoCopter,
5    RemoveHHFlightMode
6  }
```

**Listing 2: Product configuration with four delta models**

```
1  delta HexoCopter {
2
3    modify component OutputProcessor {
4      add port out Integer engine5;
5      add port out Integer engine6;
6    }
7
8    modify component FlightController {
9      modify component scp(engineCount=6);
10     add port out Integer engine5;
11     add port out Integer engine6;
12   }
13
14   modify component SteeringCmdProcessor {
15     rename component quadPowerCalc
16         as hexaPowerCalc;
17   }
18 }
```

**Listing 3: Delta for six propeller engines**

model in order to specify constraints for the order of delta model application. These constraints are necessary to resolve conflicts between deltas affecting the same architectural elements. For example, redefining a parameter in the configuration of a subcomponent is only valid, if the respective parameter exists. If the corresponding modification for adding this parameter is defined in another delta, the respective delta has to be applied before the current one and, therefore, has to be included in the order constraint (see Section 5 for more details on the used validation checks).

The order constraint is defined after the keyword `after` and specifies which deltas have to be and which ones must not be applied before the current delta. In this expression, delta model names are used as boolean values. If a delta model is applied, it is represented by the value `true`, otherwise by `false`. Specifying not only a partial order, but a logical expression allows for a more precise definition of the application order. For instance, the requirement of at least or exactly one delta from a set of deltas can be expressed. Creating a delta that depends on prior deltas requires knowledge about its predecessors, e.g., for modifying elements that have been introduced by a previous delta. However, this may be supported by tools that automatically check the consistency of the newly added delta.

A product configuration is described by explicitly specifying delta models that have to be applied to generate the product. In the current version of the language, a product configuration is independent of product features. However, an integration of our approach into a feature-based development process using tools, like FeatureIDE [21] or pure::variants [30], is possible and prototypically realized for the former. A feature is associated with one ore more deltas that realize this feature. For a particular feature selection, a product configuration containing the set of associated deltas is derived and then used to generate a concrete product.

```
1  delta HeightHold
2    after PressureSensor && !HexoCopter {
3
4    modify component SteeringCmdProcessor {
5      add port in Boolean heightHoldFlag;
6      add port in Integer heightValue;
7      add component HeightComparator hc;
8      add component HeightAdaptor ha;
9      connect quadPowerCalc.powerOutput ->
10         ha.curPowerCalc;
11     connect ha.newPowerOutput -> powerOutput;
12   }
13
14   modify component FlightController {
15     add port in Boolean heightHoldFlag;
16   }
17 }
```

**Listing 4: Delta for height hold flight support mode**

*Example.*

The following example illustrates the use of Δ-MontiArc. In this example, the architectural component `FlightController` and its subcomponents, as described in Section 2, are used as core models. The multicopter software architecture is modified with Δ-MontiArc in order to (a) operate with six instead of four rotor engines, (b) provide a flight support mode which holds the height of the multicopter and (c) remove the heading hold flight mode. To achieve these three transformations, the four delta models listed in the product configuration in Listing 2 are used. The resulting architectural modifications are illustrated in Figure 2.

The `HexoCopter` delta extends the software architecture with the support of six instead of four propeller engines (see Listing 3). The connections between the new outgoing ports of the output processor and the ones of the flight controller are not defined explicitly. They are automatically created due to the `autoconnect port` statement in the `FlightController` core definition. Furthermore, the configuration of the subcomponent `scp` in the flight controller needs to be adjusted by setting configuration parameter `engineCount` to six (l. 9). Line 15 demonstrates the use of the `rename` operator.

The `HeightHold` delta adds the ports, subcomponents and connectors to provide the architecture for the desired flight support mode (see Listing 4). If this flight support mode is enabled, the height hold flag is set to `true`. The new subcomponents `HeightComparator` and `HeightAdaptor` in the component `SteeringCmdProcessor` calculate the difference between the desired and the actual height of the multicopter and adjust the calculated power for the rotor engines accordingly. Since this delta model requires the pressure sensor to be available, it can only be applied after the `PressureSensor` delta which is specified in the order constraint (l. 2). The `PressureSensor` delta adds an incoming port to the `FlightController` and a new subcompo-

```
1  delta PressureSensor {
2    modify component FlightController {
3      add port in PressureSensorStat;
4      add component PressureEval pEval;
5    }
6  }
```

**Listing 5: Delta for adding a pressure sensor**

```
1  delta RemoveHHFlightMode {
2
3    expand autoconnect;
4
5    modify component FlightController {
6      disconnect steeringMode ->
7          scp.steeringMode;
8    }
9
10   modify component SteeringCmdProcessor {
11     disconnect steeringMode ->
12         quadPowerCalc.steeringMode;
13   }
14
15   remove unreachables;
16   introduce autoconnect port;
17 }
```

**Listing 6: Delta for removing the heading hold flight mode**

```
1  grammar DeltaMontiArc extends MontiArc {
2
3    external Constraint;
4
5    MADeltaModel =
6      "delta" Name ("after" Constraint)?
7      deltaBody:MADeltaBody;
8    //...
9    AddArcElementStatement implements Statement =
10     "add" ArcElement;
11   //...
12 }
```

**Listing 7: Example for reusing languages via inheritance and embedding**

nent (`PressureEval`) (Listing 5). Since the `HexoCopter` delta renames the subcomponent `quadPowerCalc`, it has to be applied after the `HeightHold` delta to ensure the validity of the `connect` statement in line 9. This order constraint is expressed by negating the application of the `HexoCopter` delta (Listing 4, l. 2).

Listing 6 shows the `RemoveHHFlightMode` delta which removes the heading hold flight mode. Since only one flight mode (accelerometer mode) remains, the input for setting a flight mode can be removed as well. There are basically two ways for removing unused connectors and ports. As in this scenario all connectors to be removed are only implicitly modeled with the `autoconnect` statement, it is sufficient to remove only the ports which are not used anymore. The second way is demonstrated by the delta model in Listing 6. The `expand autoconnect` operation results in all connectors being represented explicitly (l. 3). The undesired connectors can then be removed with the `disconnect` operator (l. 6f, 11f). Thereby, several ports are no longer used and can be removed by the `remove unreachables` operator (l. 15). The `introduce autoconnect port` statement removes the connectors unnecessarily created by the `expand autoconnect` statement and reestablishes the `autoconnect` concept (l. 16).

## 4. IMPLEMENTATION OF Δ-MontiArc

The language Δ-MontiArc is implemented with the MontiCore framework [12]. One benefit of MontiCore is the possibility to reuse already defined languages and their tools via inheritance and embedding. Δ-MontiArc uses both concepts in order to extend MontiArc. Listing 7 shows a simplified partial grammar definition of Δ-MontiArc to illustrate language reuse.

Δ-MontiArc inherits from the MontiArc language which is specified by the `extends` keyword, similar to programming languages such as Java (l. 1). Thus, grammar productions defined in MontiArc can also be used in Δ-MontiArc's grammar definition. This is exemplified by the `AddArcElementStatement` production which uses the `ArcElement` production of MontiArc. In the same way, the definition of several productions in the Δ-MontiArc grammar is simplified. Embedding an already defined language is realized by the keyword `external` (l. 3). The defined gaps can be used in any production, as it is the case in line 6. Here, a language for logical expressions can be integrated which is used for defining the application order constraints. The concrete language that is employed to fill the `Constraint` gap can be defined elsewhere.

A further benefit of using MontiCore is the possibility to generate a parser which processes an input model expressed in the defined language and creates an abstract syntax tree (AST). The AST can be modified by further language processing workflows defined in MontiCore. When processing an input model, the language processing workflows can be invoked in any desired order. Δ-MontiArc reuses the parsing workflow of MontiCore. The workflow for product generation from a given product configuration is defined manually and explained in more detail in the following section.

## 5. PRODUCT GENERATION

The product generation workflow of Δ-MontiArc works on three different kinds of models: MontiArc core models, delta models and product configurations. The product configuration defines a set of delta models which have to be applied to the core models to generate a particular product model. The generation workflow uses the Δ-MontiArc implementation to parse the delta models and the MontiArc implementation for loading the core models. The generation workflow consists of the following four steps:

1. Load MontiArc core models from core folders and store their ASTs and symbol tables.

2. Load the delta models which are listed in the product configuration and compute the application order.

3. Process the delta models one by one in the computed order by modifying the core/intermediate ASTs and symbol tables entries.

4. Pretty print the resulting ASTs.

In the first step, all MontiArc models stored in the defined core folders are loaded, i.e., the models are parsed, their symbol tables are built up and the default set of MontiArc's context conditions is checked [13]. The resulting AST and the infrastructure for accessing the symbol tables is stored for each core model.

In the second step, the product configuration is parsed and the listed delta models are loaded. Afterwards, the application order of these delta models is determined by evaluating the application order constraints (specified with the `after` keyword). An application order constraint is a logical expression over deltas and defines which deltas have to be or must not be applied before the current delta. An already applied delta is represented by the boolean value `true`, a not yet applied delta is represented as `false`. The application order can then be calculated by using a SAT-solver.

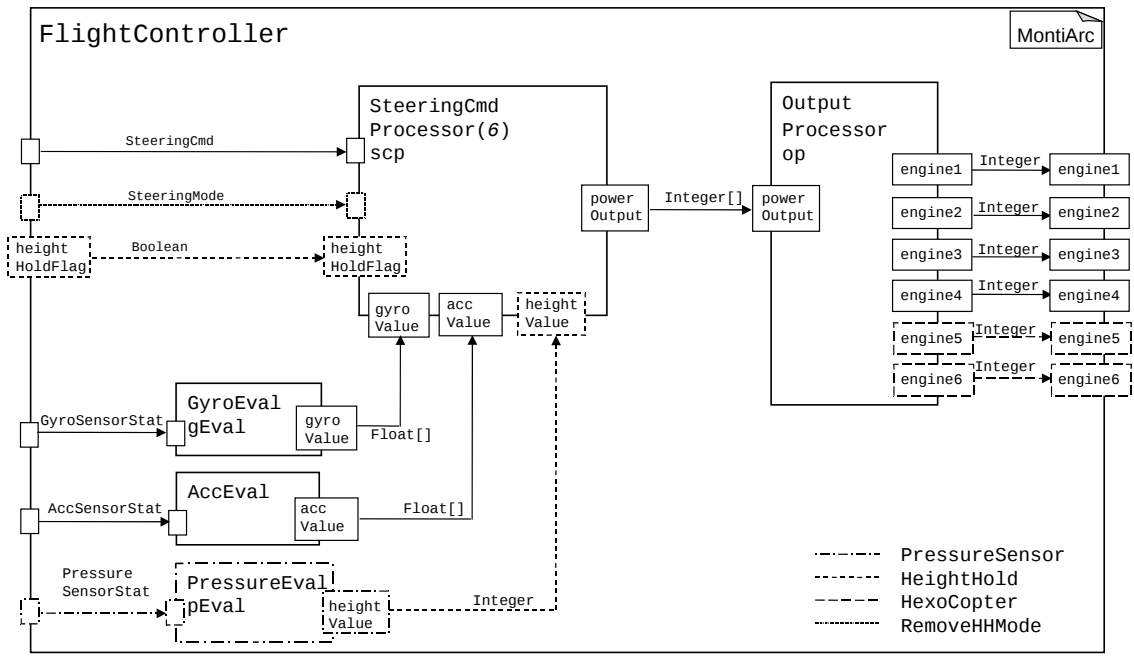

Figure 2: Flight controller after application of deltas

As an example, Figure 3 shows the resulting tree for a product configuration containing the four delta models $A$, $B$, $C$ and $D$. Delta $A$ has no defined constraint for the application order. $B$ can only be applied if $D$ has not been applied before. $C$ can be applied if either $A$ or $B$ have been applied before, but not both. $D$ can be applied after $B$ or $C$, but not after $A$. The figure shows the complete tree. One resulting complete linear order for this example is $B \rightarrow C \rightarrow D \rightarrow A$.

In the third step of the product generation workflow, the delta models are applied in the computed linear order. For each delta model, every modification operation is processed in specified order using a visitor [8] which traverses the delta model AST. For some delta modification operations, their applicability needs to be checked before execution. The applicability checks comprise the following conditions:

- A component $c$ can only be modified, if $c$ exists.

- An architectural element $ae$ must not be added to component $c$, if $c$ already contains $ae$.

- An architectural element $ae$ must not be removed from component $c$, if $c$ does not contain $ae$.

- A port $p$ must not be removed from component $c$, if $c$ contains a connector with $p$ as its source or target.

- A subcomponent $sc$ must not be removed from component $c$, if $c$ contains a connector that has a port of $sc$ as its source or target.

- An architectural element $ae_{old}$ must not be renamed to $ae_{new}$ in component $c$, if $c$ does not contain $ae_{old}$ or $c$ already contains an element of the same kind with name $ae_{new}$.

- A parameter $cp$ must not be redefined in the configuration of a subcomponent $sc$, if $sc$ does not contain a configuration parameter $cp$.

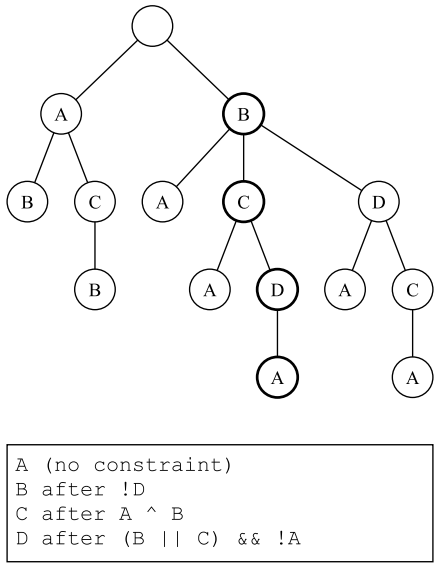

```
A (no constraint)
B after !D
C after A ^ B
D after (B || C) && !A
```

Figure 3: Tree structure for application order calculation

- A subcomponent $sc_1$ must not be replaced by a subcomponent $sc_2$ if both components have incompatible interfaces (see Section 3 for the definition of compatible interfaces).

If the currently executed delta modification operation is applicable, the AST of the modified MontiArc component is adjusted together with the corresponding symbol table entries. After the application of a delta model, the modified or newly created architectural elements are checked against a set of MontiArc context conditions. This set of context conditions comprises all checks that can be done locally on the affected elements, e.g., after adding a port the naming convention that each port name should start with a lower case letter is checked. However, not every MontiArc context condition can be checked here, since the resulting architectural model is not required to be valid after the application of a single delta model.

In the last step of the product generation workflow, the modified ASTs of all newly obtained core models are transformed back into MontiArc models. This is realized by using MontiArc's pretty printer which is implemented as a visitor that traverses the AST and prints the corresponding MontiArc statement for each node.

# 6. CASE STUDY

In order to evaluate the potential of delta-oriented variability modeling for software architectures in $\Delta$-MontiArc, we compared it in two case studies with annotative variability modeling approaches [37]. In annotative variability modeling, a model is constructed that contains all elements that are present in any possible variant. Hence, these models are often also called *150%-models* [11]. Model elements are annotated with product features and removed from the model for a product if the feature is not selected.

As first case study, we used the multicopter example presented in Section 3. For the second case study, we modeled an architecture for an anti-lock braking system (ABS) with several variants. The core architecture is used in cars and calculates individual braking pressures for four wheels by monitoring these wheels. In case of a (nearly) blocking wheel, indicated by the corresponding wheel sensor, the braking pressure for this wheel is reduced. A second variant of the ABS is an implementation for a motorcycle. This variant only has two wheels, but a separate brake for the rear wheel. The third variant is used in trucks with six wheels and includes two additional sensors for the wheels and two additional actuators to adjust the braking pressures. A traction control is supported by a fourth variant of the ABS. This variant evaluates the position of the acceleration pedal and calculates braking pressures for the wheels without traction. A fifth variant includes an electronic stability control (ESP). For the ESP, sensors measure the lateral acceleration, and specific braking pressures are calculated to avoid critical situations.

For comparing annotative and delta-oriented variability modeling, we used MontiArc as underlying ADL. For delta modeling, we used $\Delta$-MontiArc. The 150%-model architecture is realized by annotating the MontiArc elements to express their usage in specific variants. An example annotation is displayed in Listing 8 where subcomponent `scp` is configured with the value 6 in variant `HexoCopter`, and otherwise with 4. In $\Delta$-MontiArc, this is expressed by only

```
<<variant = "HexoCopter">>
component SteeringCmdProcessor(6) scp;
<<variant = "!HexoCopter">>
component SteeringCmdProcessor(4) scp;
```

Listing 8: Example for an annotated MontiArc-element

|          | MultiCopter | | ABS | |
| --- | --- | --- | --- | --- |
|          | $\Delta$-MA | 150% | $\Delta$-MA | 150% |
| LOC      | 104 | 97 | 127 | 146 |
| # Files  | 13 | 9 | 9 | 5 |
| max. LOC | 15 | 29 | 27 | 101 |
| avg. LOC | 8 | 10.78 | 14.11 | 29.2 |
| rel. VC  | 42.31% | 21.65% | 62.20% | 36.30% |

Table 2: Case study results

one `modify component` statement (see Listing 3, l. 9).

In order to evaluate both approaches with respect to their usability, we compared the size of the models by measuring the absolute lines of code (LOC) and the number of files (# Files). Both values correlate to size and complexity of a model, which is easier to comprehend if both values are low. Typically, a component is modeled in its own file. When using delta modeling, additionally each delta is also defined in its own file. As models are developed by human developers, the size of the model, in terms of LOC per file, should not exceed a certain limit to facilitate the understanding of the models. Thus, we measure the maximum lines of code (max. LOC) and the average lines of code (avg. LOC) per file which indicates understandability as smaller models are easier to read. Additionally, we measured the relative amount of variability-related code contained in the models. For the 150%-model, we divided the number of annotated elements by the total number of elements. For delta modeling, we divided the number of LOC in delta models by the total number of LOC.

The results of the evaluation are summarized in Table 2. It can be seen that the absolute number of LOC is nearly the same for both modeling approaches. However, the maximum and average number of LOC per file is higher for the 150%-approach. This means that in the annotative variability modeling approach the single files comprising parts of the model are larger since the variability referring to all possible variants has to be represented in every file together with the non-variable architectural elements. In contrast, delta modeling allows the separation of the model into a number of smaller files, reducing model complexity and increasing modularity and understandability. The relative variability-related code is nearly twice as large in the delta modeling approach than in the 150%-model. A deeper inspection of these results reveals that a complete delta is counted as variability-related code although only small portions are modification operations while larger portions refer to existing component definitions, introduce new architectural elements or define the ordering constraints. The effect of saving LOC by expressing variability using statements such as `remove unreachables` does not have consequences for the LOC contained in delta models in the considered case examples. Nevertheless, the separation of variability-related code in deltas allows identifying architectural elements contained in different variants more easily. Regarding the over-all result of the evaluation, the Δ-MontiArc approach seems to have a better scalability, in particular for larger models, than the 150%-model approach.

## 7. RELATED APPROACHES

Other approaches to represent architectural variability can be classified in two main directions [37]: annotative (or negative) and compositional (or positive) modeling approaches. Annotative approaches consider one model (that is usually non-hierarchical) representing all products of the product line. Variant annotations, e.g., using UML stereotypes [38, 10] or presence conditions [5], define which parts of the model have to be removed to derive a concrete product model. The orthogonal variability model (OVM) [29] represents variability in a model that is separated from the artifact model. The variability modeling language (VML) [24] specializes the ideas of OVM for architectural models.

Compositional approaches associate model fragments with product features that are composed for a particular feature configuration. In [18, 37, 27], models are constructed by aspect-oriented composition. Feature-oriented model-driven development (FOMDD) [34] combines feature-oriented programming (FOP) with model-driven engineering. In [7], model fragments are merged in order to provide the variability model of a product line. Apel et al. [1] apply model superposition to compose model fragments. Plastic partial components [28] represent architectural variability by extending partially defined components with variation points and associated variants. Variants can be cross- or non-cross-cutting architectural concerns that are composed with the common component architecture by weaving mechanisms. In the Koala component model [35, 36], the variability of a component is described hierarchically by the variability of its subcomponents. The selection between different subcomponent variants is realized by variation points, called switches, that are designated components. For each variation point, a particular component variant can be selected. In [15], we extended MontiArc by hierarchical variability modeling concepts similar to the Koala approach.

Apart from positive and negative variability representations, model transformations are used for capturing product variability. The common variability language (CVF) [17] represents the variability of a base model by rules describing how modeling elements of the base model have to be substituted in order to obtain a particular product model. In [20], graph transformation rules capture the variability of a single kernel model comprising all commonality. In [19], architectural variability is represented by change sets containing additions and removals of components and component connections that are applied to a base line architecture. Delta modeling [4, 32] that is applied to represent variability in Δ-MontiArc can also be classified as a transformational approach. In contrast to the above mentioned variability modeling approaches, Δ-MontiArc facilitates proactive, reactive, and extractive product line development [23] via flexible modification operations that are specified in delta models, while still allowing modular variability representations.

Other textual ADLs, similar to MontiArc, that could be extended with variability modeling concepts are Acme [9] or xADL [6]. An extension of Acme can be achieved using its property mechanism. A variation point may be embedded in a property and modeled by a plain string. This approach, however, is error prone, as these strings have to be interpreted manually. Syntax highlighting or further modeling support for variation points cannot be provided by Acme. xADL can be extended by defining new XML schemes. As human readability of XML files is poor, xADL does not match our extensibility requirements, too. As pointed out in this paper, MontiArc relying on the language extension mechanisms of MontiCore is the ideal candidate for the language extensions that have to be made for Δ-MontiArc.

Apart from MontiCore [12], several meta-case tools exist that allow the development of domain-specific (modeling) languages. ASF+SDF [2] is a meta-environment for language development that offers support for transformations, code generation, and source code analysis. It allows a modular definition of syntax, but does not reflect these concepts at the tooling level. Another framework is the Grammar Deployment Kit (GDK) [22] that supports development of grammars and language processing tools. However, concepts for language extensions, such as inheritance, are not supported. If meta-modeling techniques, like EMF [3] are used for domain-specific language development, two different descriptions for abstract and concrete syntax have to be developed and kept in line. Using MontiCore, concrete as well as abstract syntax is generated out of a single grammar. This way possible inconsistency and redundancy problems are avoided.

## 8. CONCLUSION

Δ-MontiArc as presented in this paper is an extension of a previous proof-of-concept version [14] with specific modification operations to support delta-oriented architectural variability modeling. We have shown an efficient and fast language development process for this variability modeling language that reuses language-defining artifacts as well as the language processing infrastructure of the underlying ADL based on the DSL framework MontiCore. Δ-MontiArc is compared with annotative variability modeling approaches to show that delta models are less complex and easier to understand.

For future work, we are planning to carry out larger case studies in order to evaluate the scalability of Δ-MontiArc. Furthermore, we plan to develop a delta extension of AJava [16], an architectural programming language that itself is an extension of MontiArc. In this way, the efficient development of product lines of distributed systems can be supported by a seamless delta-oriented development process from architecture to implementation is feasible.